\documentclass[journal,draftcls,onecolumn,12pt,twoside]{IEEEtran}%

\usepackage{amsfonts}
\usepackage{amsmath}
\usepackage{amssymb}
\usepackage{graphicx}
\usepackage{dsfont}
\usepackage{algorithm}
\usepackage{algorithmic}
\usepackage{multirow}
\usepackage{slashbox}
\usepackage{enumerate}
\usepackage{subfigure}
\usepackage{stfloats}
\usepackage{caption}
\usepackage{subfigure}
\usepackage{stfloats}

\graphicspath{{Figures/}}
\DeclareGraphicsExtensions{.eps}

\normalsize

\begin{document}
%
\title{ARUM: Polar Coded HARQ Scheme based on Incremental Channel Polarization}
%
%
%

\author{Kai~Chen, 
        Liangming~Wu,
        Changlong~Xu,
        Jian~Li,
        ~Hao~Xu,
        and~Jing~Jiang
\thanks{The authors are with the Wireless Lab of Qualcomm Co., Ltd (e-mail: kaichen@ieee.org)}%
}


\maketitle

\begin{abstract}

A hybrid ARQ (HARQ) scheme for polar code, which is called active-bit relocation under masks (ARUM), is proposed.
In each transmission, the data bits are encoded and bit-wisely XOR-masked using a binary vector before being transmitted through the channel.
The masking process combines multiple transmissions together which forms another step of inter-transmission channel transform.
The reliabilities are updated after every transmission, and the less reliable bits in earlier ones are \emph{relocated} to the more reliable positions at the latest transmitted block.
ARUM is a very flexible HARQ scheme which allows each transmission to have a different mother code length and to adopt independent rate-matching scheme with sufficient channel state feedback in HARQ process.
Simulation shows that ARUM can obtain near-optimal coding gain.

\end{abstract}


%
\IEEEpeerreviewmaketitle

\section{Introduction}
\label{sec:intro}

%

\IEEEPARstart{P}{olar} codes, invented by Ar{\i}kan \cite{ORIGIN}, are proved to be capacity-achieving under successive cancellation (SC) decoder over binary-input discrete memoryless channels (B-DMCs).
For finite-length cases, it is found that concatenated with cyclic redundancy check (CRC) codes, polar code under SC list (SCL) decoding achieves competitive performance to  turbo codes or low-density parity-check codes (despite that the decoding complexity of polar associated with list decoding is subject to further studies to make it practically appealing) \cite{SCLTAL}\cite{imdec_TCOM}\cite{CAdec}\cite{acadec}.
Therefore, polar code is a competitive candidate in future communication systems and has been adopted as the coding scheme for control channels for the 5G cellular system.

In wireless communication systems, hybrid automatic repeat request (HARQ) technique is widely used to obtain a higher spectrum efficiency.
After encoding, the transmitter sends only part of the coded bits into the channel; when a decoding failure signal is fed back from receiver, another part of the coded bits are incrementally transmitted.
Polar coded HARQ scheme with incremental freezing (IF) have been studied in \cite{IRIF}, where a part of less reliable bits in earlier transmissions are encoded and retransmitted using a lower rate polar code. Receiver first decodes the latest received code block and marks them as known bits when decoding the former received blocks.
IF is proven to be a capacity-achieving rate-less coding scheme over time invariant channels, but because the channel independencies between every transmissions are not fully utilized, its performance with finite code length and over fading channels should be further improved.

In this paper, a novel HARQ scheme of polar code called ARUM is proposed as an improved version of IF.
At each transmission, the data bits are encoded and are bit-wisely XOR-masked by a binary vector before sent into the channel.
The data bits for each retransmission are duplicated from the selected less reliable active bits, and the mask vector is generated according to the previous transmissions.
The masking process combines multiple transmissions together which forms another step of channel transform.
Similar to IF, part of the less reliable bits are relocated to the more reliable positions in the latest transmission.
Because of the polarization effect of the additional channel transform, ARUM can achieve a better finite-length performance than IF.

Another inspiring work is \cite{HW}, where the channel transform matrix of polar code is extended to be double sized at each retransmission.
By jointly decoding the multiple transmissions, it can achieve the same performance as a directly generated polar code.
However, just as pointed out in \cite{HW}, it requires the length of retransmitting blocks to grow as multiples of $2$, and it is inconvenient to perform a flexible rate-matching for every transmission.
ARUM, by contrast, is a very flexible scheme, it allows each transmission to have a different code length and to adopt independent rate-matching scheme.

The remainder of the paper is organized as follows.
\mbox{Section \ref{section_background}} briefly reviews the basic idea of polar coding and the rate-matching schemes.
\mbox{Section \ref{section_arum}} describes the construction, encoding and decoding of the proposed scheme.
\mbox{Section \ref{section_simulation}} evaluates the performance through simulations.
Finally, \mbox{Section \ref{section_conclusion}} concludes the paper.

\section{Polar Coding and its Rate-Matching Schemes}
\label{section_background}

\subsection{Polar Coding}

Let $W$ denote a BMC channel with channel transition probability $W\left(y|x\right)$,
After channel polarization transform on $N$ independent uses of $W$, we obtain $N$ synthesized binary-input channels $W^{(i)}_{N}$, $i=1,2,\cdots,N$ with transition probabilities
\begin{equation}
\label{equ_polarized_channels}
     {W}^{(i)}_{N}(\mathbf{y}, \mathbf{u}_{[1:i-1]}|\mathbf{u}_{[i]})=\sum\limits_{u_{[i+1:N]}}{\frac{1}{2^{N-1}} \prod\limits_{{k}=1}^{N}{W(\mathbf{y}_{[k]}|\mathbf{x}_{[k]})}}
\end{equation}
\noindent{where} $\mathbf{x}=\mathbf{u} \cdot {\mathbf{G}_{N}}$, and
the channel transform matrix ${\mathbf{G}_{N}}={\mathbf{B}_{N}} \cdot \mathbf{F}^{\otimes n}$, in which ${\mathbf{B}_{N}}$ is the $N \times N$ bit-reversal permutation matrix and the kernel matrix of Ar{\i}kan's polar code is
\begin{equation}
\label{eqn:kernel_F}
{\mathbf{F}}=\left[ \begin{matrix}   1 & 0  \\   1 & 1  \\ \end{matrix} \right]
\end{equation}
where $\mathbf{F}^{\otimes n}$ denotes the $n$-th Kronecker power of $\mathbf{F}$ and code length $N=2^n$, $n \in \{1,2,3,...\}$.

After this channel transform, $K$ of the synthesized channels ${W}_N^{(i)}$ with larger reliabilities, with index set denoted as $\mathcal{A}$, are used to carry data bits, and the rest are fed with frozen zero bits.
In this way, we get an $(N,K)$ polar code.
To estimate the reliabilities, Gaussian approximation (GA) \cite{GA} is an efficient and accurate method under binary-input AWGN channel.

\subsection{Rate-Matching Schemes of Polar Codes}

In practical application, to transmit $K$ data bits, the required code length $M$ will not always be some power of $2$.
Rate-matching is used to adjust the length of mother code from $N$ to the system required code length $M$.
In 3GPP standard \cite{NR38212}, the rate-matching schemes are classified into three categories: repetition, puncturing and shortening.

Repetition, which is applied when $N \le M$, forms an $M$-length sequence by duplicating $M-N$ of the coded bits and appending them to the $N$-length codeword of the mother code.
Receiver collects and sums up the log-likelihood (LLR) values corresponding to the identical coded bits, and feeds the resulting $N$-length LLR vector to the decoder.
The LLR of some bit $b$ is defined as
\begin{equation}
\label{eqn_llr}
LLR(b) = \log {\frac{\Pr\{b=0\}}{\Pr\{b=1\}}}
\end{equation}

Puncturing \cite{QUP} is applied when $N>M$, where $N-M$ bits of the $N$-length codeword are not transmitted through channel.
Receiver sets the LLR values of these punctured bits to zeros.

Shortening is a special case of puncturing \cite{SHORTEN}.
By utilizing the code structure, $N-M$ synthesized channels are selected to carrying zero-bits which makes $N-M$ of the coded bits to be fixed zeros.
Then, $K$ data bits are transmitted through the most reliable ones of the rest $M$ channels.
In this case, it is unnecessary to transmit these $N-M$ zero-valued coded bits, and the receiver can directly set the corresponding LLRs to $+\infty$ when decoding.

The selection of the repeated or punctured positions has strong impact on finite-length performance \cite{PARALLEL}.
The best known universal solution is performing a bit-reversal permutation on the coded bit sequence, then repeating/puncturing $|M-N|$ bits sequentially from the first, or in a reverse order from the last.
Particularly, for punctured polar code, this solution is known as quasi-uniform puncturing (QUP) .

The channel reliabilities are changed after rate-matching, so the optimal active set $\mathcal{A}$ should be adjusted accordingly \cite{PARALLEL}.
However, in HARQ procedure, there is no chance to change the set $\mathcal{A}$ once it is determined at the first transmission.
Therefore, these existing rate-matching schemes cannot be directly applied in HARQ transmissions.

\section{The Proposed Polar Coded HARQ Scheme}
\label{section_arum}

\subsection{Active-bit Relocation Under Masks}

\begin{figure}[!t]
\centering{
\includegraphics[width=0.85\columnwidth]{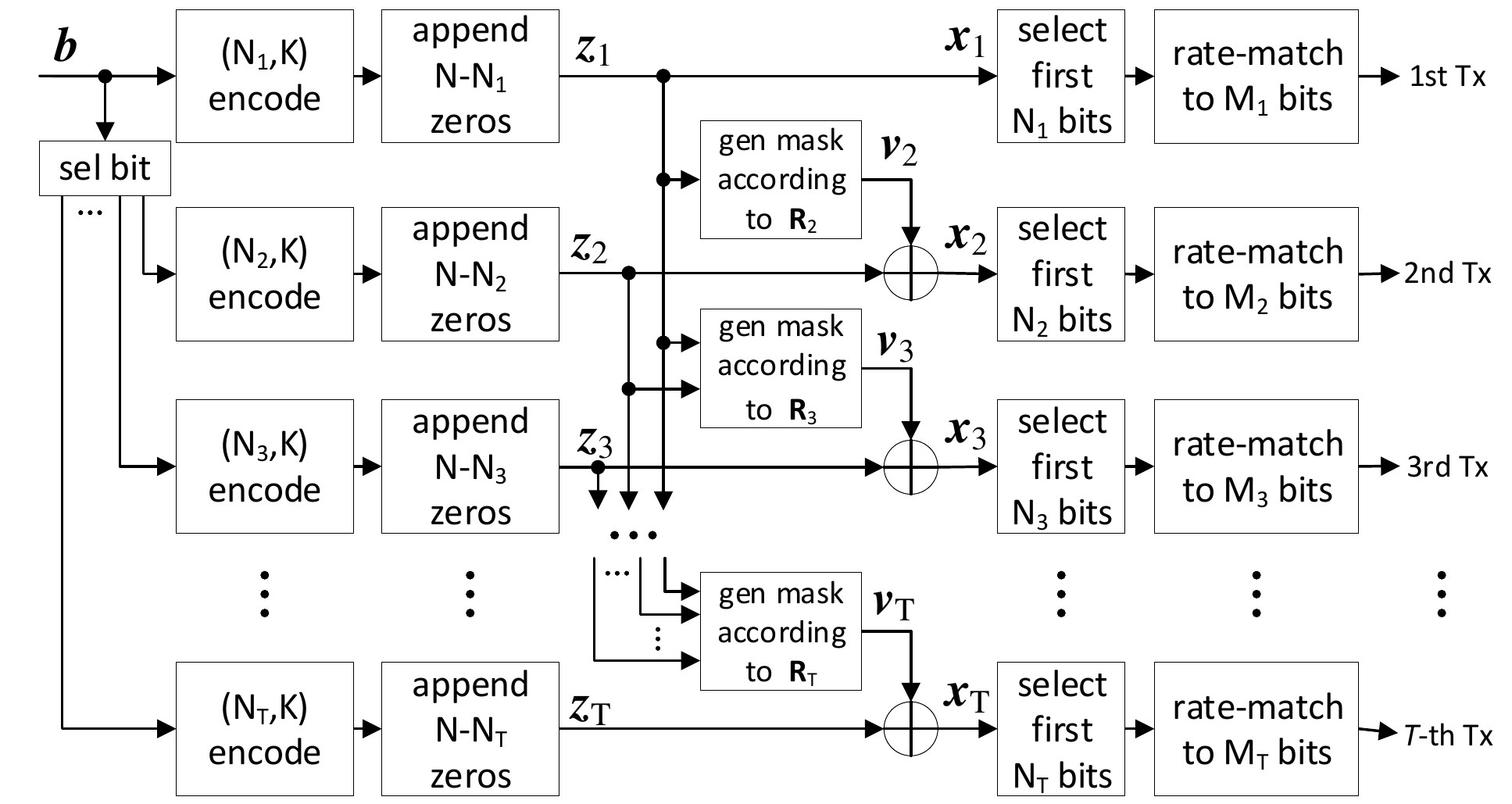}}
\caption{The proposed HARQ scheme with ARUM.}
\label{fig:arum_scheme}
\end{figure}

\begin{figure}[!t]
\centering{
\includegraphics[width=0.85\columnwidth]{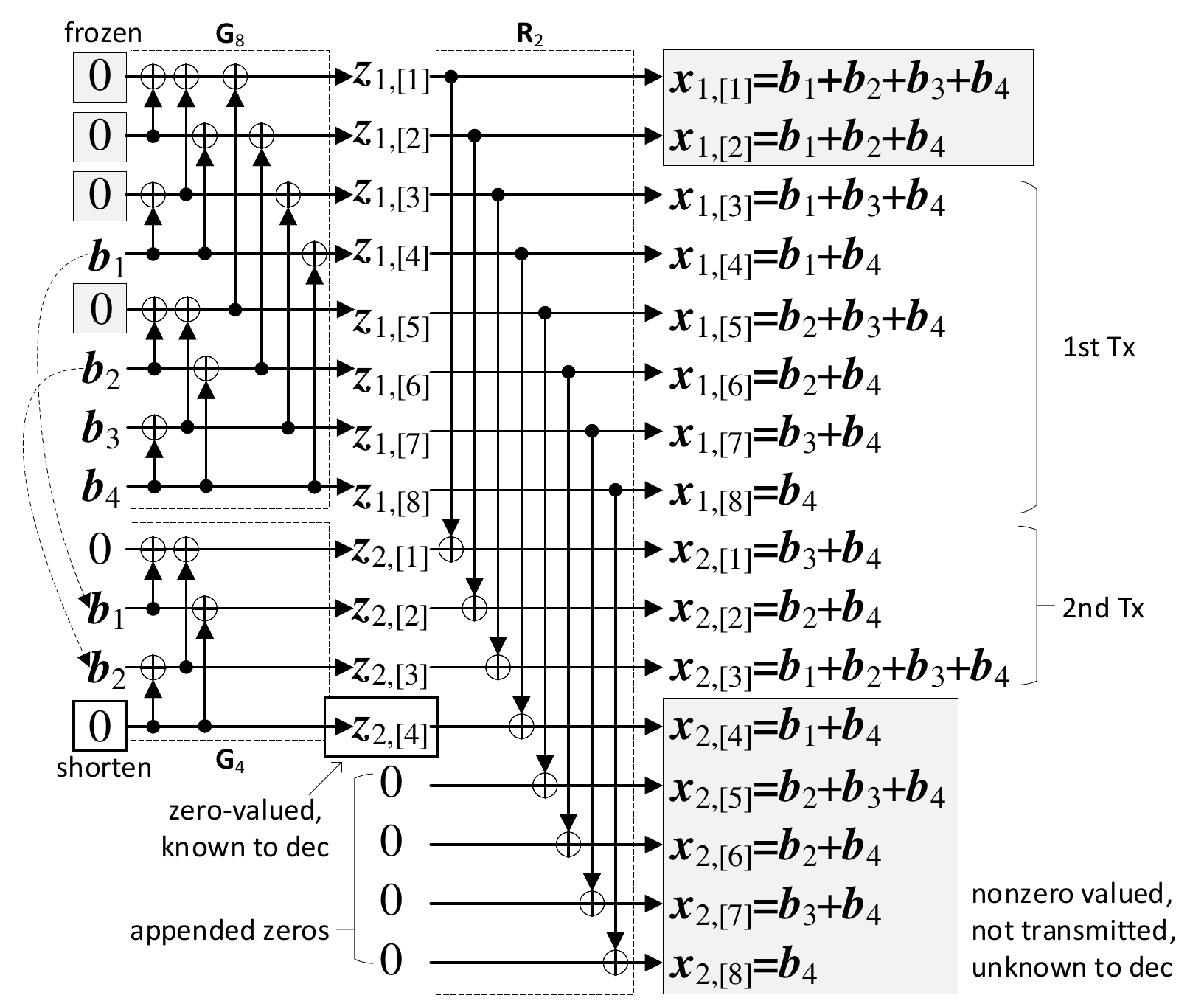}}
\caption{A toy example of two transmissions.}
\label{fig:example}
\end{figure}

The proposed HARQ scheme is shown in Fig.\ref{fig:arum_scheme}.

The mother code length $N_t$ for the $t$-th transmission is determined by rate-matching algorithms, $t=1,2,\cdots, T$.
To achieve a potential better performance, the selection of rate-matching scheme could be jointly optimized with ARUM.
However, this topic is beyond the scope of this paper and will be left for future study.

At the $t$-th transmission, part of the data bits $\mathbf{b}$ are selected and encoded into an $N_t$-length codeword.
For the first transmission, all the $K$ bits of data sequence $\mathbf{b}$ are transmitted.
After that, $N-N_t$ zeros are appended to the end of the result codeword and an $N$-length vector $\mathbf{z}_t$ is obtained.
When $t>1$, an $N$-length mask vector $\mathbf{v}_t$ is generated based on $\mathbf{R}_t$.
The $i$-th element of mask $\mathbf{v}_t$ is obtained as follows:
\begin{equation}
\label{eqn:mask_gen}
\mathbf{v}_{t, {[i]}} = \left(\mathbf{z}_{1,{[i]}}, \mathbf{z}_{1,{[i]}},\cdots, \mathbf{z}_{t-1,{[i]}}\right) \cdot \mathbf{R}_{t, [1:t-1, t]}
\end{equation}
where $i \in \{1, 2, \cdots, N\}$, $\mathbf{R}_t$ is a $t \times t$ matrix, and $\mathbf{R}_{t, [1:t-1, t]}$ denotes a column vector which consists of elements in the first $(t-1)$ rows of the $t$-th column of $\mathbf{R}_t$.
Then, the vector $\mathbf{v}_t$ is bit-wise XOR masked over $\mathbf{z}_t$ to obtain $\mathbf{x}_t$.
Finally, the first $N_t$ bits of $\mathbf{x}_t$, i.e, $\mathbf{x}_{[1:N_t]}$, are fed into rate-matching module to obtain the $M_t$-length sequence for the $t$-th transmission.

The data bits at the less reliable positions in previous $(t-1)$ transmissions are duplicated and sent at a set of more reliable ones in the $t$-th block.
The decoder combines all the $t$ received LLR vectors and performs inter-block successive cancellation decoding from the latest received block to the first received block.
After decoding the $t$-th block, part of the data bits in block $1$ to $(t-1)$ which have been duplicated in the $t$-th transmission are marked as non-active data bits.
When decoder encounters a non-active data bit, it takes the same operations as it is a frozen bit, except that it could be nonzero valued: the decoder simply makes a hard decision on the non-active bit based on previous decoding result and updates the path metric accordingly.
Therefore, the error probability of some data bit $\mathbf{b}_k$, $k \in \{1,2,\cdots, K\}$, is dominated by the error rate of its copies in the latest transmitted block, w.l.o.g., the $t$-th block.
In this sense, this data bit $\mathbf{b}_k$ which is carried by the $1$st block is \emph{relocated} to the $t$-th block.

At the receiver, to decode the $t$-th block, the LLRs of multiple transmissions are collected and combined according to $\mathbf{R}_t$.
The LLRs of shortened bits are set to $+ \infty$, and decoding this $(N_t, K_t)$ polar code using SC/SCL algorithm, where $K_t$ is the number of active bits in the $t$-th transmission.
To help decode the rest blocks $t-1,...,1$, the decoded active bits in block $t$ are stored, and the signs of LLR of $t$-th blocks are flipped according to its re-encoded codeword.
Note that, if SCL is adopted for joint decoding, a list of candidate active bits, along with a list of corresponding path metrics, should be passed down to decode the rest blocks.
%

A toy example of two transmissions are given in Fig. \ref{fig:example}.
At the first transmission, the rate matching scheme is puncturing from $N_1=8$ to $M_1=6$ bits.
Four data bit $(\mathbf{b}_1,\mathbf{b}_2,\mathbf{b}_3,\mathbf{b}_4)$ are encoded as a half-rate polar code,
where the first two coded bits are punctured and the rest are transmitted.
At the second transmission, the rate matching scheme is to construct a shortened code with $M_2=3$ from a mother code with $N_2=4$.
The two less reliable bits $\mathbf{b}_1$ and $\mathbf{b}_2$ are encoded and masked by $\mathbf{z}_1$, and the first three masked coded bits are transmitted.
When joint decoding over these two transmissions, the LLRs of the untransmitted bits are set to zeros.
Note that, when decoding the $2$nd block, LLR of $\mathbf{z}_{2,[4]}$ should be set to $+\infty$ because it is related to a shortened bit with a known value zero;
but LLR of $\mathbf{x}_{2,[4]}$ should be set to zero, because the corresponding bit in mask vector $\mathbf{x}_{1,[4]}$ could be nonzero.

\subsection{Two-Step Incremental Channel Polarization Transform}

\begin{figure}[!t]
\centering{
\includegraphics[width=0.8\columnwidth]{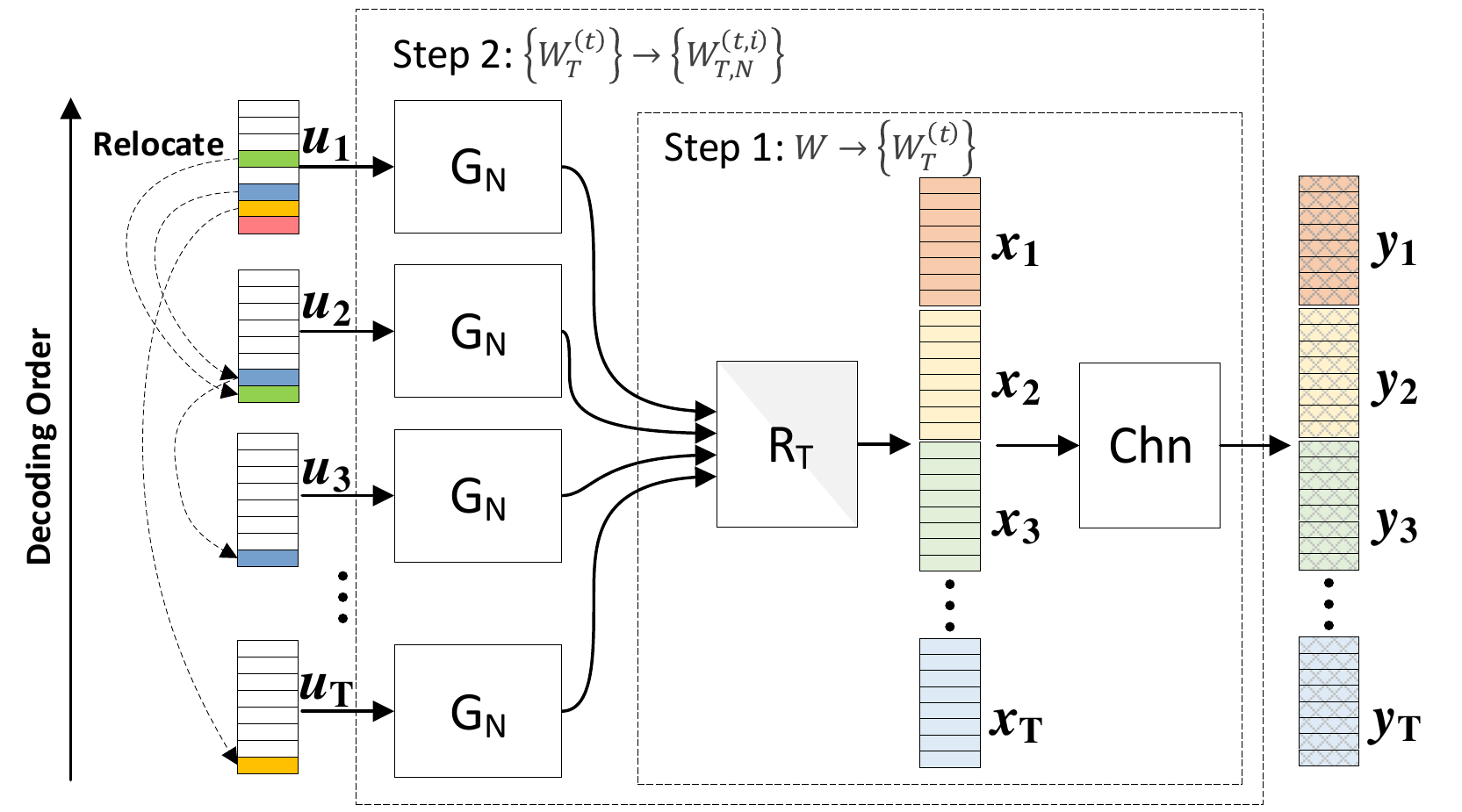}}
\caption{Two-step channel transform of ARUM.}
\label{fig:channel_transform}
\end{figure}

As the demonstrated in Fig. \ref{fig:channel_transform}, the ARUM transmissions can be translated into polar coding over incremental channel polarization, which consists of two steps of channel transform.

The first step combines $T$ channel uses of underlying channel $W$ in $T$ transmissions, and transforms them into a set of synthesized channels $\left\{ W_T^{(t)} \right\}$, $t=1,2,\cdots,T$, with transition probabilities
\begin{eqnarray}
\label{eqn:step1}
W_T^{(t)}\left(y_1, y_2, \cdots, y_T, z_{t+1}, z_{t+2}, \cdots, z_{T} | z_t \right) \quad \quad && \nonumber \\
\quad \quad \quad = \sum_{z_{1}, \cdots, z_{t-1}} {\left(\frac{1}{2^{T-1}} \cdot \prod_{k=1}^{T}{W\left(y_k|x_k\right)} \right)} &&
\end{eqnarray}
where
\begin{equation}
\left( x_1,x_2,\cdots,x_T \right)=\left( z_1,z_2,\cdots,z_T \right) \cdot \mathbf{R}_T
\end{equation}
Note that the transform matrix $\mathbf{R}_T$ should be designed to make the decoding order to be from the $T$-th transmission to the first, so $\mathbf{R}_T$ must be an upper-triangular matrix.
Moreover, because block $1$ to $t$ are transmitted before the construction of block $t$, the corresponding coded bit should be aligned with those in previous transmissions.
Therefore, the matrix $\left\{\mathbf{R_t}\right\}$ for each transmission $t=1,2,\cdots,T$, should share a nested structure,
i.e., for any $s > t$, $\mathbf{R}_t$ is the sub-matrix which consists of the first $t$ rows and first $t$ columns of matrix $\mathbf{R}_s$,
\begin{equation}
\label{eqn:R_property}
\mathbf{R}_t = \mathbf{R}_{s, [1:t, 1:t]}
\end{equation}

\newcounter{TempEqCnt}
\setcounter{TempEqCnt}{\value{equation}}
\newcounter{eqn_num_step2}
\setcounter{eqn_num_step2}{7}
\newcounter{eqn_num_after_step2}
\setcounter{eqn_num_after_step2}{8}
\setcounter{equation}{\value{eqn_num_step2}}

\begin{figure*}[!t]
\begin{small}
\begin{equation}
\label{eqn:step2}
W_{T,N}^{(t,i)}\left(\mathbf{y}_1,\cdots,\mathbf{y}_T,\mathbf{u}_{t+1},\cdots,\mathbf{u}_{T},\mathbf{u}_{t,[1:i-1]}|\mathbf{u}_{t,[i]}\right)
= \sum_{ \begin{smallmatrix} \mathbf{u}_{1}, \cdots, \mathbf{u}_{t-1} \\
\mathbf{u}_{t, [i+1:N]} \end{smallmatrix}
} {\prod_{j=1}^{N}\left(\frac{{W_T^{(t)}}\left(\mathbf{y}_{1,[j]},\cdots,\mathbf{y}_{T,[j]},\mathbf{z}_{t+1,[j]},\cdots,\mathbf{z}_{T,[j]},|\mathbf{z}_{t,[j]}\right)}{2^{N-1}} \right)}
\end{equation}
\end{small}
\hrulefill
\end{figure*}

\setcounter{equation}{\value{TempEqCnt}}

For each $t\in \{1,2,\cdots,T\}$, a second step of channel transform, which is the same as conventional polar coding, is applied to $N$ independent uses of channel $W_T^{(t)}$.
The set of resulting channels are denoted by $\left\{ W_{T,N}^{(t,i)} \right\}$, $i=1,2,\cdots,N$, with channel transition probabilities given in (\ref{eqn:step2}).

\setcounter{equation}{\value{eqn_num_after_step2}}

At the $T$-th transmission, after two-step channel transform, the most reliable $K$ synthesized channels are selected for carrying active data bits, where the channel index set is
\begin{equation}
\mathcal{A}_T = \left\{a_k | k=tN+i, t\in\{1,\cdots, T\}, i\in\{1,\cdots, N\} \right\}
\end{equation}
with size $\left|\mathcal{A}_T\right|=K$.
The bits with indices in difference set $\mathcal{A}_{T-1}-\mathcal{A}_{T}$ are \emph{relocated} to positions indexed by $\mathcal{A}_{T} - \mathcal{A}_{T-1}$.

The choice of matrix $\mathbf{R}_t$ of the first step channel transform plays an important role in ARUM.
Similar to the conventional polar code, $\mathbf{R}_t$ can be built by Kronecker products of transpose of kernel matrix (\ref{eqn:kernel_F}),
\begin{equation}
\label{eqn:kernel_arikan}
\mathbf{R}_{t,\text{A}} = \mathbf{R}_{S,\left[1:t,1:t\right]}
\end{equation}
where
\begin{equation}
\mathbf{R}_S =
\left[
    \begin{matrix}
        1 & 1 \\
        0 & 1 \\
    \end{matrix}
\right] ^{\otimes \left \lceil \log_2{t}\right \rceil}
\end{equation}
With this Ar{\i}kan kernel $\mathbf{R}_{t,\text{A}}$, joint decoding over received LLR vectors of $t$ transmissions can be equivalent to decode a conventional polar code with code length $tN$, in most cases.
One exception is that when shortening rate-matching modes are adopted for $t>1$, as the example shown in Fig.\ref{fig:example}.

If the $\mathbf{R}_t$ is designed to be an identity matrix, the mask vectors $\{v_t\}$ are always all-zero vectors. In this case, ARUM is degraded to incremental freezing (IF) studied in \cite{IRIF}.
\begin{equation}
\label{eqn:kernel_if}
\mathbf{R}_{t,\text{IF}} =
    \left[
        \begin{matrix}
        1 &   &   &  \\
          & 1 &   &  \\
          &   & \ddots & \\
          &   &   &  1\\
        \end{matrix}
    \right]_{t \times t}
\end{equation}
As introduced in Section \ref{sec:intro}, ARUM under IF kernel (\ref{eqn:kernel_if}) is capacity-achieving but suffers performance loss under finite code length.
ARUM with Ar{\i}kan kernel (\ref{eqn:kernel_arikan}) can well polarize the channel, but it has growing column weights when $t \ge 4$, which requires more complex combinations of the receiving vectors from multiple transmissions.

In this paper, first-xor-latest (FL) masking is proposed for ARUM, which provides a better tradeoff between the complexity and performance.
With FL masking, the mask vector is always set to be $\mathbf{v}_t = \mathbf{z}_1$, so the transform matrix is
\begin{equation}
\label{eqn:kernel_FL}
\mathbf{R}_{t,\text{FL}}=\left[
        \begin{matrix}
        1 & 1 & 1 & \cdots & 1 \\
          & 1 & 0 & \cdots & 0 \\
          &   & \ddots & \ddots &\vdots  \\
          &   &        & 1& 0 \\
          &   &        & & 1 \\
        \end{matrix}
    \right]_{t \times t}
\end{equation}
The first three transmissions under FL masking are exactly the same as that under Ar{\i}kan kernel masking (\ref{eqn:kernel_arikan}).
From the \mbox{$4$th} block, ARUM under FL masking has a less complexity.

\section{Simulation Results}
\label{section_simulation}

The proposed polar coded HARQ scheme with ARUM is simulated over BPSK modulated AWGN channels.
W.o.l.g, the rate-matching scheme for each transmission follows the 3GPP standard \cite{NR38212} with parameters determined by the data length of the first transmission $K$ and the number of available transmitting bits for each transmission $M_t$, respectively.
The data bits consists of $K-16$ information bits and $16$ CRC bits.
The block error rate (BLER) after $T$ ARUM transmissions is compared with that of $(M_1+M_2+\cdots+M_T, K)$ polar code constructed following 3GPP standard or GA with QUP rule.

\begin{figure}[!t]
\centering{
\includegraphics[width=0.82\columnwidth]{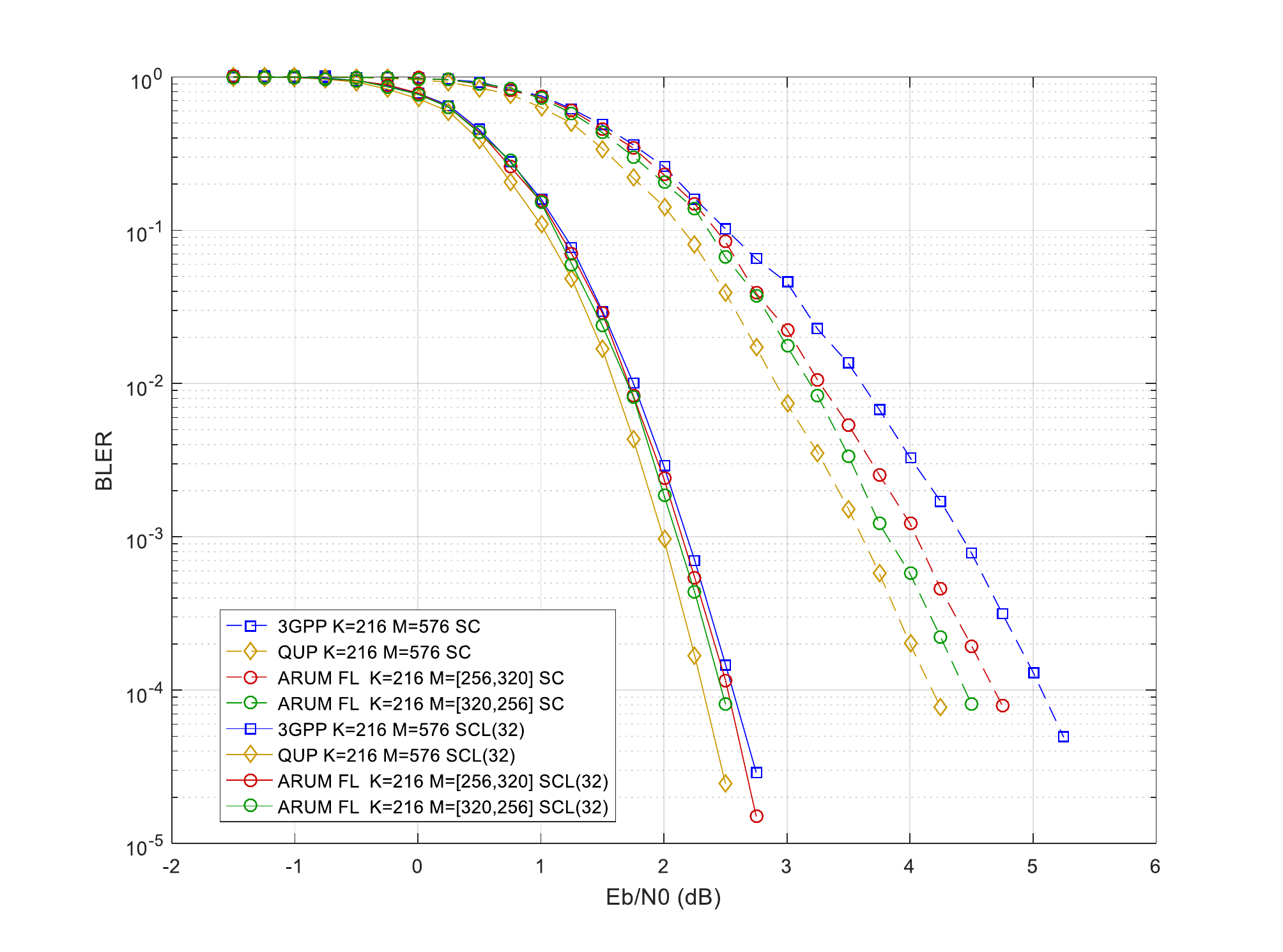}}
\caption{Joint decoding of two ARUM transmissions.}
\label{fig:sim2tx}
\end{figure}

\begin{figure}[!t]
\centering{
\includegraphics[width=0.82\columnwidth]{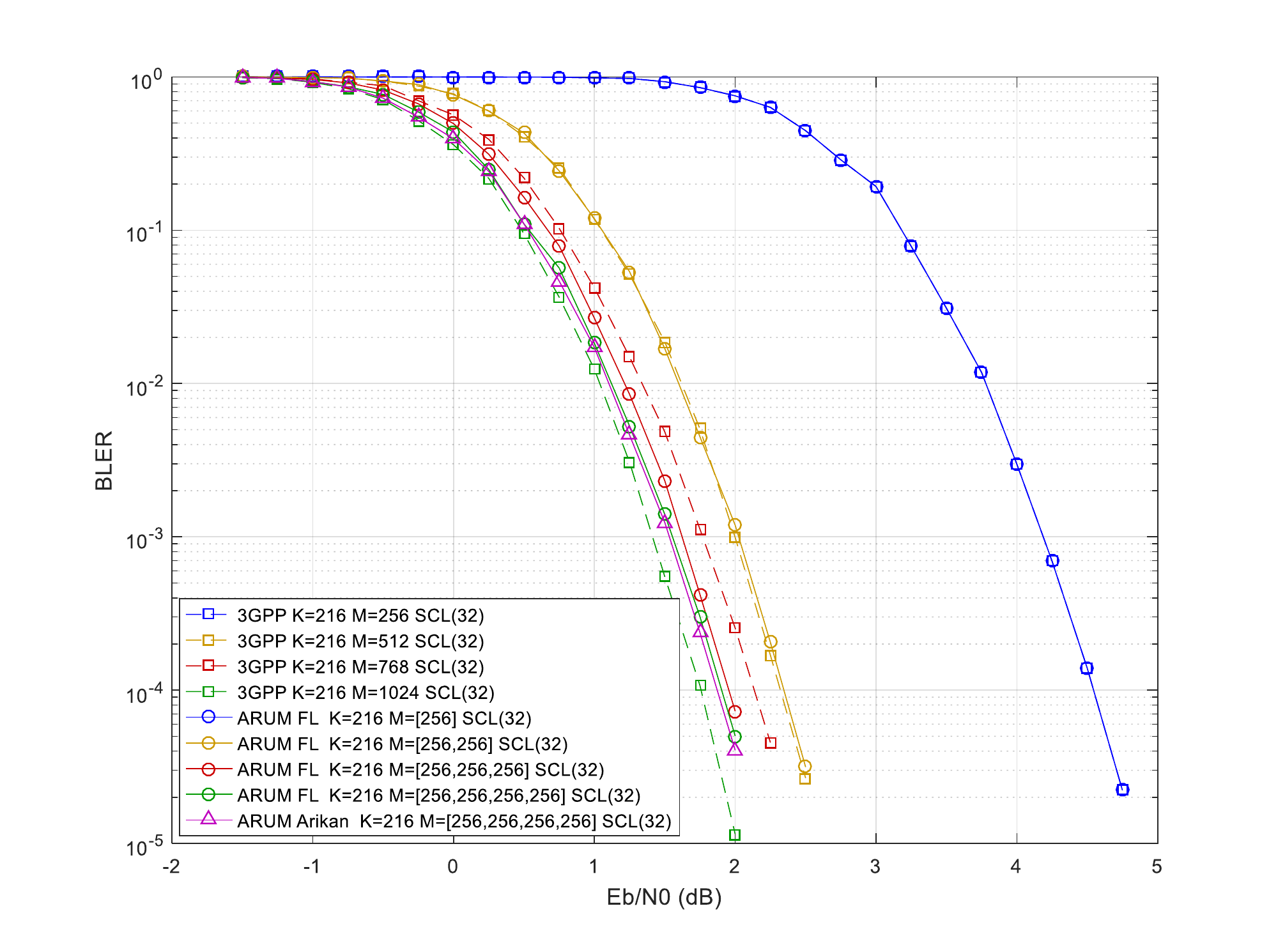}}
\caption{Joint decoding of up to four ARUM transmissions.}
\label{fig:sim4tx}
\end{figure}

In Fig. \ref{fig:sim2tx}, two ARUM transmission cases are simulated: one is with $M_1=256 < M_2=320$; and the other is with $M_1=320 > M_2=256$.
When $M_t=320$, $t=1,2$, the corresponding rate-matching scheme works in shortening mode with mother code length $N_t=512$.
After joint decoding over two ARUM transmissions, both cases have comparable performances against a (576, 216) polar code which is directly constructed following 3GPP standard or QUP rule.

BLER curves up to four ARUM transmissions are given in Fig. \ref{fig:sim4tx}, where every transmission has the same code length $M_t=256$.
Performance after one/two transmissions well match the curves of the $256$/$512$-length polar codes.
Performance after three transmissions is even slightly better than that of $768$-length 3GPP code, because the 3GPP code under puncturing has to trade off between the code description complexity and the performance.
Performance after four transmissions is slightly worse than that of directly constructed code. Because the relocation can only de-active the active bits in former transmissions, but cannot work in a reverse direction, the number of active bits left in the 3rd block is less than that under a direct construction.
Nevertheless, the loss is less than $0.2$dB and could be smaller when the code rate of the 1st transmission is lower.
The performance of four ARUM transmissions under Ar{\i}kan kernel masking (\ref{eqn:kernel_arikan}) is also provided.
However, its improvement over FL kernel is trivial, while it requires more complicated combinations when combining the four LLR vectors.

\section{Conclusion}
\label{section_conclusion}
A flexible polar coded HARQ scheme called ARUM is proposed, which allows the code lengths and rate-matching scheme of every transmission to be different from each other.
ARUM scheme is constructed based on a two-step channel polarization transform, where one of the two steps is the same as conventional polar codes and the other is an incremental channel transform that performed inter-transmissions.
First-xor-latest masking is proposed and recommended for the incremental channel transform, which provides a better trade-off between complexity and performance under ARUM.
The simulations shows that by joint decoding multiple transmissions, ARUM can obtain near-optimal coding gain, i.e., the performance is very close to that of a directly constructed polar code with the target code length.

However, the channel condition and rate-matching scheme is time-varying in practical system. Using GA to evaluate the reliablilities of synthesized channel (\ref{eqn:step2}) is still too complex for online computation. Future work will develop a more efficient online construction method of ARUM.


\ifCLASSOPTIONcaptionsoff
  \newpage
\fi


\begin{thebibliography}{1}

\bibitem{ORIGIN}
E.~Arikan, ``Channel polarization: a method for constructing capacity-achieving codes for symmetric binary-input memoryless channels,'' \emph{IEEE Trans. Inf. Theory}, vol. 55, no. 7, pp. 3051-3073, Jul. 2009.

\bibitem{SCLTAL}
I.~Tal and A.~Vardy, ``List decoding of polar codes,'' \emph{IEEE Int. Symp. Inform. Theory (ISIT)}, pp. 1-5, 2011.

\bibitem{imdec_TCOM}
K. Chen, K. Niu, and J. Lin, ``Improved Successive Cancellation Decoding of Polar Codes,'' \emph{IEEE Trans. Commun.}, vol. 61, no. 8, pp. 3100-3107, Aug. 2013.

\bibitem{CAdec}
K. Niu and K. Chen, ``CRC-aided decoding of polar codes,'' \emph{IEEE Commun. Lett.}, vol. 16, no. 10, pp. 1668-1671, Oct. 2012.


\bibitem{acadec}
Y. Fan, C. Xia, J. Chen, C. Tsui, et.al, ``A Low-Latency List Successive-Cancellation Decoding Implementation for Polar Codes'', \emph{IEEE Journal on Selected Areas in Commun.}, vol. 34, no. 2, pp. 303-317, 2016.


\bibitem{IRIF}
B. Li, D. Tse, K. Chen, and H. Shen, ``Capacity-achieving rateless polar codes,'' \emph{IEEE Int. Symp. Inform. Theory (ISIT)}, pp. 46-50, Aug. 2016.

\bibitem{HW}
L. Ma, J. Xiong, and Y. Wei, ``An Incremental Redundancy HARQ Scheme for Polar Code,'' arXiv:1708.09679v1, Aug 2017.

\bibitem{GA}
P. Trifonov, ``Efficient design and decoding of polar codes,'' \emph{IEEE Trans. Commun.}, vol. 60, no. 11, pp. 3221-3227, Nov. 2012.

\bibitem{NR38212}
\emph{Multiplexing and channel coding}, 3GPP TS38.212 V15.0.0, Dec 2017.

\bibitem{QUP}
K. Niu, K. Chen, and J. Lin, ``Beyond turbo codes: Rate-compatible punctured polar codes'', in Proc. IEEE ICC, Budapest, Hungary, 2013

\bibitem{SHORTEN}
R. Wang and R. Liu, "A novel puncturing scheme for polar codes", IEEE Commun, Letters, vol. 18, pp. 2081-2084, Dec. 2014

\bibitem{PARALLEL}
K. Chen, K. Niu, and J. Lin, ``Practical polar code construction over parallel channels,'' IET Communications, vol.7, no.7, pp.620-627, 2013.


\end{thebibliography}
\end{document}